\documentstyle[twoside,fleqn,espcrc2,amsfonts]{article}

\newcommand{\eq}[1]{\begin{equation}#1\end{equation}}              
\newcommand{\eqs}[1]{\begin{eqnarray}#1\end{eqnarray}}
\newcommand{\eqlab}[1]{\label{eqn:#1}}
\newcommand{\eqref}[1]{(\ref{eqn:#1})}
\newcommand{\Eqref}[1]{eq.~(\ref{eqn:#1})}

\newcommand{\seclab}[1]{\label{sec:#1}}
\newcommand{\secref}[1]{\ref{sec:#1}}
\newcommand{\Secref}[1]{sect.~\ref{sec:#1}}
\newcommand{\Secsref}[1]{Sects.~\ref{sec:#1}}

\newcommand{\RR}{{\Bbb R}}
\newcommand{\CC}{{\Bbb C}}
\newcommand{\ZZ}{{\Bbb Z}}

\newcommand{\pd}{\partial}
\newcommand{\tr}{{\rm tr}\,}
\newcommand{\ord}{{\rm Ord}\,}
\newcommand{\psdo}{\Psi}
\newcommand{\Tr}{{\rm Tr\,}}
\newcommand{\Trc}{{\rm Tr}_{\rm c}}
\newcommand{\res}{{\rm Res}\,}
\newcommand{\sym}{{\rm Sym}\,}
\newcommand{\hilb}{{\cal H}} 			
\newcommand{\hpl}{\hilb_+}			
\newcommand{\hmi}{\hilb_-}			
			
\newcommand{\sop}{\varepsilon}			
\newcommand{\fock}{{\cal F}(\hilb)} 		
\newcommand{\fvac}{|0\rangle}			
\newcommand{\bop}{{\cal B(\hilb)}} 		
\newcommand{\ham}{H}				
\newcommand{\nm}{|\hskip-.6pt|}			
\newcommand{\crn}{a^*_n}
\newcommand{\crm}{a^*_m}

\newcommand{\ann}{a_n}
\newcommand{\anm}{a_m}

\newcommand{\Dsl}{\not\!\!\pd }
\newcommand{\crt}{\tilde{c}_{\rm R}}
\newcommand{\dom}{\Omega_d}
\newcommand{\oalgt}{{\cal L}_3}
\newcommand{\oalgth}{\hat{\cal L}_3}
\newcommand{\ohom}{\rho}
\newcommand{\shom}{\bar{\rho}}
\newcommand{\idl}{{\cal I}^{-3}}

\newcommand{\lieg}{\frak g}
\newcommand{\Map}[2]{{\rm Map}(#1\!,#2)}
\newcommand{\mapS}{\Map{S^1}{\lieg}}
\newcommand{\mapM}{\Map{M}{\lieg}}
\newcommand{\mapMd}{\Map{M_d\,}{\lieg}}

\newcommand{\lgen}[1]{{\cal L}_{#1}(\hilb)} 	
\newcommand{\lo}{\lgen{1}}
\newcommand{\lt}{\lgen{2}}
\newcommand{\lp}{\lgen{p}}

\newcommand{\glgen}[1]{{\frak gl}_{#1}(\hilb)} 	
\newcommand{\glo}{\glgen{1}}
\newcommand{\glt}{\glgen{2}}
\newcommand{\glp}{\glgen{q}}

\newcommand{\gloh}{\widehat{\glo}}		
\newcommand{\glth}{\widehat{\glt}}

\newcommand{\gglgen}[1]{{\frak gl}_{#1}} 
\newcommand{\gglo}{\gglgen{1}}

\newcommand{\gglp}{\gglgen{q}}
\newcommand{\glopsdo}{\psdo\gglo}

\newcommand{\ggloh}{\widehat{\gglo}}

\newcommand{\wh}[1]{\widehat{#1}}
\newcommand{\dwh}[1]{\widehat{\widehat{#1}}}

\newcommand{\AmS}{{\protect\the\textfont2
  A\kern-.1667em\lower.5ex\hbox{M}\kern-.125emS}}

\title{\vbox{\vskip-60pt\hbox to\hsize{\normalsize G\"oteborg-ITP-96-15}
		\vskip-8pt
		\hbox to\hsize{{\normalsize\tt 
hep-th/9612167}}\vskip 35pt}
On higher-dimensional loop algebras, pseudodifferential operators \\
	and Fock space realizations} 

\author{Anders Westerberg\address{Institute of Theoretical Physics, \\
        Chalmers University of Technology and G\"oteborg University, \\ 
        S-412 96 G\"oteborg, Sweden}
}
       
\begin{document}

\begin{abstract}
We discuss a previously discovered \cite{CFNW-I} extension of the
infinite-dimensional Lie algebras $\mapM$ which generalizes the Kac--Moody 
algebras in 1+1 dimensions and the Mickelsson--Faddeev algebras in 3+1 
dimensions to manifolds $M$ of general dimensions.  
Furthermore, we review the method of regularizing current algebras in
higher dimensions using pseudodifferential operator (PSDO) symbol
calculus. 
In particular, we discuss the issue of Lie algebra cohomology of PSDOs and
its relation to the Schwinger terms arising in the quantization process.  
Finally, we apply this regularization method to the algebra of
ref.~\cite{CFNW-I} with partial success, and discuss the remaining 
obstacles to the construction of a Fock space representation.

\end{abstract}


\maketitle

\section{Introduction}

The importance of infinite-dimensional Lie algebras and their representation 
theory for two-dimensional quantum field theory has become overwhelmingly
clear over the past two decades. A prime example of such algebras are, 
of course, the affine Kac--Moody (KM)  algebras, which are realized in 
physical models as extensions of loop algebras $\mapS$ or as  two-dimensional 
current algebras, and as such allow for  generalizations to higher dimensional
systems.

However, the usefulness of these higher-dimensional analogues is hampered 
by their poorly developed representation theory; generalizing the techniques 
and methods used in two dimensions has turned out to be a very difficult task.
The most extensively studied case in this respect is probably the 
Mickelsson--Faddeev (MF) algebras \cite{Fa-84,Mi-85}, which arise as anomalous 
constraint algebras in chiral gauge theory.   

Here we shall describe an attempt at obtaining interesting 
representations of a particular extension of the algebras $\mapMd$ of maps 
from a $d$-dimensional manifold $M_d$ to a finite-dimensional Lie algebra
$\lieg$ 
that was discovered \cite{CFNW-I} in the context of $p$-branes. 
The primary reason for the difficulties encountered in the study of
higher-dimensional current algebras is the additional regularization needed
to obtain one-particle Hilbert space operators that by
a standard normal-ordering procedure can be elevated to Fock space operators
furnishing a unitary highest weight (HW) representation of the algebra.
Following Mickelsson \cite{Mi-I}, we will employ PSDO symbol calculus to 
perform this regularization.  
After a presentation of the algebra in \Secref{ploop} we review some 
necessary backgrund material on fermionic Fock space constructions of 
current algebras and on PSDOs in \Secsref{glalgs}
and \secref{psdo}, respectively. In \Secref{app} we present our 
main results, and finally end the paper by some comments in \Secref{concl}.  
  
\section{Loop algebras in general dimensions}
\seclab{ploop}
In ref.~\cite{CFNW-I} we presented the following extension
of $\mapMd$:
\eqs{
\lefteqn{[\![(X;z), (Y;w)]\!]} \cr
 &&=([X,Y]+[dX,dY];\,\alpha_d\!\!\int_{M_d}\!\!\tr X\,dY)
\eqlab{mapext}
}
Here $X\!=\!\bigoplus_{i=0}^{[d/2]}X^{(2i)}$ and 
$Y\!=\!\bigoplus_{i=0}^{[d/2]}Y^{(2i)}$
are sums of differential forms of even degree taking values in the 
compact Lie algebra $\lieg$, while $z,w\in\CC$ and 
$\alpha_d$ is a normalization constant.
The brackets on the right hand side are ordinary (not graded)
commutators. Furthermore, wedge products between forms are understood and 
forms of degree higher than $d={\rm dim}\,M_d$ are assumed to vanish. 
The integral, which defines the central term of the algebra, selects the 
terms in the form expansion proportional to the volume form. 

The characteristic features of these algebras are that their formulation
is diffeomorphism invariant, valid for general $d$ and reduces to 
the well-known KM and MF algebras for $d=1,3$, respectively. 
Furthermore, in contrast to the MF algebras in general dimensions \cite{Mi-85},
the algebras \eqref{mapext} are always linear. 
We should also mention that \eqref{mapext} defines a non-abelian
extension of $\mapMd$ when $d\geq4$ \cite{CFNW-I}.  
  
The algebra \eqref{mapext} may alternatively be formulated as an algebra
of smeared current densities:
\eqs{
[T(X),T(Y)] \!\!\!&=&\!\!\! T([X,Y]+ [dX,dY]) \nonumber\\
 &&\quad + \;k\,\alpha_d\!\!\int_{M_d}\!\!\tr X\,dY
\eqlab{smearalg}
}
In this form, the algebra was used in ref.~\cite{CFNW-I} as the 
starting-point of 
a $p$-loop space formulation of a model for odd-dimensional $p$-branes 
coupled to 
a background Yang--Mills field; by imposing \eqref{smearalg} as the algebra 
obeyed by certain functional operators appearing in the BRST transformations, 
it was found to be possible to construct a nilpotent BRST operator, as well
as a $p$-loop space gauge covariant derivative and 
the corresponding curvature tensor. This construction was an attempt
to get round the difficulties caused by non-linearities arising in a
corresponding formulation \cite{DD-92,Di-93} based on a Kaluza--Klein type 
$p$-brane action \cite{DDS-92}. 
Although, so far, we have not been able to derive our algebra from any 
particular $p$-brane action, recent
developments in the subject of Dirichlet $p$-branes (D-branes) gives
reason for some optimism in this respect. 
For instance, similar geometrical constructions involving sums of 
differential forms of various degrees have been found very useful
in the construction of effective world-volume actions for 
D-branes \cite{GHT,CvGNW,CvGNSW,APS,BT}. 

\section{Abstract current algebras}
\seclab{glalgs}

We now wish to look for unitary HW representations of the 
above algebra on a fermionic Fock space. To this end, let us
first review the general framework for the study of current 
algebras that we will make use of \cite{Lu-76,CR-87}.

Thus, consider a one-particle description of free Dirac or Weyl fermions 
$\psi\in\hilb={\rm L}^2(\RR^d)\otimes V=\hpl\oplus\hmi$, where $\hpl$
($\hmi$) denotes the positive (negative) energy subspace as determined
by the hamiltonian $H$, and $V$ is a finite-dimensional vector
space carrying the internal (spin and ``color'') degrees of freedom.  
This quantum mechanical model is second-quantized by introducing
a Fock vacuum $\fvac\in\fock$ such that 
\eq{
a_n^*\fvac=0,\;n<0, \qquad a_n\fvac=0,\;n\geq0, 
}
where the creation and annihilation operators obey the canonical
anticommutation relations
\eq{
\{\crm,\ann\}=\delta_{mn},\;\{\crm,\crn\}=\{\anm,\ann\}=0.
}
The second-quantized observables $\hat{A}$ acting on $\fock$ are defined by 
the linear map
\eq{
A\longmapsto\hat{A}=\sum_{m,n}A_{mn}:\crm\ann:,
\eqlab{sqmap}
}
with the normal-ordering
\eq{
:\crm\ann:\,= \left\{\begin{array}{ll}
			-\ann\crm &	\mbox{if $m,n<0$} \\
			\phantom{-}\crm\ann &	\mbox{otherwise}
			\end{array} \right.
}
defined to give $\hat{A}$ a vanishing vacuum expectation value. 

In order for $\hat{A}$ to be a well-defined operator on $\fock$ it has to 
obey the crucial condition 
\eq{
\nm \hat{A}\fvac\nm<\infty,
}
which may readily be reformulated as $A\in\glo$, where for $q\in\ZZ_+$
the general linear algebras of $\hilb$ are defined as
\eq{
\glp=\left\{A\in\bop\,|\, [\sop,A]\in\lgen{2q}\right\}.
\eqlab{glpdef}
}
Here $\sop={\rm sign}\,\ham$, $\bop$ is the space of bounded operators
on $\hilb$, and the Schatten ideals $\lo\subset\lt\subset...$ 
are defined by
\eq{
\lp=\{A\in\bop\;|\; \Tr |A^*A|^{p/2}<\infty\},
}
with $\Tr\!$ denoting the Hilbert space trace. 
In particular, operators in $\lo$ and $\lt$ are called trace-class and 
Hilbert--Schmidt (HS), respectively.

As it turns out, the Fock space operators obtained by 
second-quantizing $\glo$ do not satisfy the same Lie algebra as their 
one-particle ancestors. Instead one finds the commutation relations
\eq{
[\hat{A},\hat{B}]=\widehat{[A,B]}+c_1(A,B),
\eqlab{glohat}
}
defining the universal central extension $\gloh$ of $\glo$. 
The Schwinger term
\eq{
c_1(A,B)={1\over4} \Tr \sop[\sop,A][\sop,B]
\eqlab{lundb}
}
of this current algebra is a non-trivial Lie algebra two-cocycle originally 
found by Lundberg \cite{Lu-76}\footnote{
This cocycle, which we will hence refer to as the Lundberg cocycle,
is perhaps more commonly known as the Kac--Peterson cocycle \cite{KP-81}.}.

Since $\gloh$ has unitary HW representations \cite{KP-81}, 
a basic strategy for constructing such representations of various 
extensions of $\mapMd$ is to embed the latter in $\gloh$.
On the one-particle level there is a natural embedding 
$\mapMd\hookrightarrow\bop$ as multiplicative operators:
\eq{
X\mapsto \tilde{X}:\; \tilde{X}\psi(x)=X(x)\psi(x),\;x\in M_d.
}
However, due to ultra-violet divergencies such operators belong to $\glo$ 
iff $d=1$; in general $\mapMd\hookrightarrow\glp$ only for $q>d/2$.
For $M_d=S^1$ one obtains in this way spinor representations of 
affine KM algebras \cite{KP-81}.  
When $d>1$ on the other hand, normal-ordering is not sufficient to yield 
well-defined second-quantized operators. 
One method that has been employed to deal with this problem is to enlarge
the Fock space by considering current algebras based on $\glp$ for $q>1$ 
\cite{MR-88,FT-90,La-94}. However, it has been shown \cite{Pi-89}
that $\glth$, which is the relevant algebra for $d=3$, has no unitary HW
representations.

An alternative strategy is to perform additional regularization of the
one-particle observables before second-quantizing. For the 
MF algebra in $d=3$ this approach leads to projective 
representations in terms of one-cocycles valued in the space of functionals 
of the gauge fields \cite{Mi-90}. A convenient way to perform the 
regularization \cite{Mi-I} is by means of PSDO symbol calculus, since the UV 
behavior can then be characterized and treated very explicitly. 
In \Secref{app} we will apply this method to the algebra discussed in 
\Secref{ploop}, but first we will give the prerequisites from the theory 
of PSDOs. 
    
\section{PSDOs and their cohomology}
\seclab{psdo}

Consider the Hilbert space $\hilb= {\rm L}^2(\dom)\otimes\CC^N$ of 
square-integrable vector-valued functions with support on some compact
domain $\dom\subset\RR^d$
A pseudodifferential operator $S$ on $\hilb$ is defined via its
symbol $s:\dom\times\RR^d\rightarrow{\frak gl}(N,\CC)$ by
\eq{
S\psi(x)={1\over{(2\pi)^d}}\int s(x,p)\tilde{\psi}(p)e^{ix\cdot p}d^d p,
}
where $\tilde{\psi}(p)=\int\psi(x)e^{-ix\cdot p}d^d x$ is the Fourier transform
of $\psi\in\hilb$. 
Here we shall consider only smooth symbols
of integral order, the order $\ord(s)=m$ of a symbol $s$ being
defined in terms of its leading asymptotic behavior 
$s(x,p)={\cal O}(|p|^m)$ for large $|p|$. 
We denote the space of such
symbols as $\sym^m$ and the corresponding space of PSDOs as $\psdo^m$.
In particular, $S$ is called
an infinitely smoothing operator ($S\in\psdo^{-\infty}$) if its symbol 
vanishes more rapidly than any power of $p$ as $|p|\rightarrow\infty$. 
Furthermore, two PSDOs $S$ and $S'$ are called
equivalent, denoted as $S\approx S'$ (or $s\approx s'$ for their 
symbols), if $S-S'\in\psdo^{-\infty}$. Up to equivalence, a PSDO 
$S\in\psdo^m$ with symbol $s$ is determined by an asymptotic 
expansion
\eq{
s(x,p)\approx\sum_{k\leq m}s_k(x,p),
}
where each term $s_k$ is taken to be smooth and homogeneous of degree $k$ 
in $p$ outside some finite radius.

Composition of PSDOs is defined on the equivalence classes 
$\sym^\infty/\sym^{-\infty}$ by the star product
\eq{
\sigma(S)*\sigma(S')\approx\sigma(SS'),
}
where $S\in\psdo^m$, $S'\in\psdo^n$ and $\sigma:\psdo^m\rightarrow\sym^m$ 
denotes the symbol map. The asymptotic behavior of the star product 
$s*s'\in\sym^{m+n}$ of $s\in\sym^m$ and $s'\in\sym^n$ is 
\eq{
(s* s')(x,p)\!\approx\!\!\sum_{k=0}^{\infty}\!{{(-i)^k}\over{k!}}
{{\pd ^k s}\over{\pd  p_{i_1}...\pd  p_{i_k}}}{{\pd ^k s'}\over{\pd  x^{i_1}...
 \pd  x^{i_k}}}.}

A PSDO $S$ is bounded iff $S\in\psdo^0$ and belongs to the Schatten
ideal $\lp$ iff $S\in\psdo^m$ with $m<-d/p$. The latter result is easily
established by taking into account the asymptotic behavior  
${\cal O}(|q|^m)$ and using the expression\footnote{
Here $\tr\!$ denotes the finite-dimensional matrix trace.}
\eq{
\Tr S=\int_{\dom\times\RR^d}\tr s(x,p)\,d^d x\,d^d p
}
for the Hilbert space trace of a PSDO $S\in\psdo^k$ with symbol $s$.
Hence, $S$ is trace-class iff $\ord S<-d$ and Hilbert--Schmidt iff
$\ord S<-d/2$. A property of $\Tr\!$ that will be of central importance
below, is that it is not well defined on the equivalence classes
of trace-class PSDOs. This follows
immediately from the observation that $\Tr\!$ does not vanish on 
$\psdo^{-\infty}$.
Moreover, we can notice that by calculating the finite-dimensional trace and
the integral in the proper order the class of PSDOs which yield a finite
result can be extended; more specifically, one can define a conditional
trace \cite{LM-94} by
\eq{
\eqlab{condtr}
\Trc S=\lim_{\Lambda\rightarrow\infty}\int_{|p|\leq\Lambda}{{d^d p}
\over{(2\pi)^d}}\int_{\dom}\!d^dx\,\tr s(x,p),
}
with the integrations performed in the order indicated. Operators for which
\eqref{condtr} converges are called conditionally trace-class.
Of course, for trace-class PSDOs $\Tr\!$ and $\Trc$ coincide.

In contrast to $\Tr\!$ and $\Trc$, the Wodzicki resi\-due \cite{Wo-87,Gu-85}
\eq{
\res(s)={1\over(2\pi)^d}\int_{|p|=1}\!\!\!\!d\Omega_p\int_{\dom}\!\!\!d^d x\,
\tr s_{-d}(x,p)
}
gives a well-defined (and unique) trace-functional on 
$\psdo^\infty/\psdo^{-\infty}$.
Here $s_{-d}$ is the coefficient of order $-d$ in an asymptotic expansion
for $s$.
The Wodzicki residue can in turn be used to define a non-trivial two-cocycle 
on the Lie algebra of PSDOs known as the Radul cocycle \cite{Ra-91}:
\eq{
c_{\rm R}(S,S')=-{1\over2}\res([\log |p|,s]_* * s').
}
Here $[.,.]_*$ denotes the star commutator defined in the obvious way.

So far the choice of Hilbert space $\hilb$ has been fairly arbitrary. To
make contact with the framework of \Secref{glalgs} we will from this point on
consider Weyl fermions in $d=2n-1$ dimensions with $\hilb={\rm L}^2(\dom)
\otimes\CC^{n-1}_{{\rm spin}}\otimes\CC^N_{\rm color}$. 
Equipped with the grading operator $\sop={\rm sign}\,\ham$, where 
$\ham=-i\!\!\Dsl$ is the
free Weyl Hamiltonian, we can embed the PSDOs in the general linear algebras
$\gglp$. Recalling the definition \eqref{glpdef} and the Schatten ideal 
conditions above we find, in particular, that a PSDO $S$ belongs to $\gglo$, 
which we write as $S\in\glopsdo$, iff $\ord[\sop,S]<-d/2\;$\footnote{We  will 
refer to this as the Hilbert--Schmidt condition.}. 
In terms of the corresponding symbols this reads
\eq{
S\in\glopsdo\Leftrightarrow\ord[\sop,s]_*<-{d\over2},
}
where we have used the notation $\sop$ also for the symbol of the 
grading operator.
When expanding the star commutator, this condition translates into equations
relating the coefficients of an asymptotic expansion for $s$ which can
be solved order by order, thus yielding PSDOs with a good second-quantized
behavior.

Turning to the second-quantization, one immediately has to face the
problem that the Lundberg cocycle $c_1$ entering in the algebra $\ggloh$
of the second-quantized operators is not well defined on the equivalence
classes $\glopsdo/\psdo^{-\infty}$ of second-quantizable PSDOs due to the fact 
that it is defined in terms of the Hilbert space trace $\Tr$\footnote{ 
Actually, it turns out that the proper trace to use when 
defining the Schatten ideals, and hence also the Lundberg cocycle, is not 
$\Tr\!$ but the conditional trace which in general may be 
defined as
\eq{
\Trc(A)={1\over2}\Tr(A+\sop A\sop).
}
}.
Nevertheless, we have shown in \cite{CFNW-II} that the restriction 
of the Lundberg cocycle to $\glopsdo$, defined in terms of the conditional 
trace \eqref{condtr}, is cohomologically equivalent to a ``twisted'' 
version \cite{Mi-94}
\eq{
\crt(S,S')=-{1\over2}\res(\sop*[\log|p|,s]_* * s')
\eqlab{trad}
}
of the Radul cocycle. In other words, for any two operators 
$S,S'\in\glopsdo/\psdo^{-\infty}$, defined by their asymptotic 
expansions,
\eq{
c_1(S,S')-\crt(S,S')=\delta\lambda(S,S'),
\eqlab{cohorel}
}
where $\delta$ is the Lie algebra coboundary operator and $\lambda$ is a 
certain one-cochain on $\glopsdo/\psdo^{-\infty}$.
For further details and discussion the reader is referred to 
ref.~\cite{CFNW-II}.

Notice that this is a rather general result independent of any
particular algebra which one might try to embed in $\glopsdo/\psdo^{-\infty}$.
Next we shall use it in the second-quantization of the 
algebra \eqref{mapext}.

\section{Fock space realization}
\seclab{app}
Finally, we are ready to consider the issue of Fock space 
realizations of the algebras \eqref{mapext}. 
We shall consider the case $M_d={\bf R}^3$, as it turns out that this
is the highest dimension in which the regularization method works 
(at least partially)
for the particular algebras \eqref{mapext}. Preliminary results on this
construction were reported in ref.~\cite{Fe-94}.

Define the maps $\ohom:\oalgt\rightarrow\glopsdo$ and 
$\shom=\sigma\circ\ohom$, where $\sigma$ is the symbol map,
by the asymptotic expansions
\eqs{
\eqlab{rhozero}
\shom(X^{(0)})\!\!\!\!\!&=&\!\!\!\!\!X^{(0)}\!+\!S^i{{\pd X^{(0)}}
 \over{\pd x^i}}\!+\!
 {1\over2}S^{ij}{{\pd ^2 X^{(0)}}\over{\pd x^i\pd x^j}}\!+\!\cdots\!\!\!\!\\
\eqlab{rhotwo}
\shom(X^{(2)})\!\!\!\!\!&=&\!\!\!\!\!{1\over2} A^{ij} X^{(2)}_{ij}\!+\!
{1\over6}A^{ijk}\pd _i X^{(2)}_{jk}\!+\!\cdots\!\!\!\!
}
Here $S^i$ is of order $-1$, $S^{ij}$ and $A^{ij}$ of order $-2$, etc.
Moreover, we have introduced the notation $\oalgt$ for the Lie algebra
\eq{
[\![X,Y]\!]=[X,Y]+[dX,dY].
}
Its central extension $\oalgth$ defined by the two-cocycle 
\eq{
\omega_3(X,Y)= -{1\over6\pi}{1\over(2\pi)^3}\int_{\RR^3}\tr X\,dY
}
is then the algebra \eqref{mapext} under consideration.

We impose two conditions on the map $\shom$:
\eqs{
\eqlab{rhohs}
\ord[\sop,\shom(X)]_* \!\!\!&\leq\!\!\!& -2, \\
\eqlab{rhohom}
[\shom(X),\shom(Y)]_* \!\!\!&=&\!\!\! \shom([X,Y]+[dX,dY]).
}
The first one is the Hilbert--Schmidt condition ensuring that $\ohom(X)\in
\glopsdo$; the second is the homomorphism condition.
Inserting \eqref{rhozero} in \eqref{rhohs} determines $S^i$. Using also
\eqref{rhotwo} and \eqref{rhohom}, $S^{ij}$ and $A^{ij}$ can then be 
expressed in terms of $S^i$. Explicitly, we find 
($\sop=p\!\!\!/\,/|p|=p^i\sigma_i/|p|$)
\eqs{
S^i\!\!\!\!&=&\!\!\!\! {i\over2}\sop{{\pd\sop}\over{\pd p_i}} 
	={1\over2p^2}\epsilon^{ijk}p_j\sigma_k, \nonumber\\
S^{ij} \!\!\!\!&=&\!\!\!\! {1\over2}(S^i S^j+S^j S^i)-{i\over2}
	({\pd S^i\over\pd p_j}+{\pd S^j\over\pd p_i}), \\
A^{ij} \!\!\!\!&=&\!\!\!\! {1\over2}(S^i S^j-S^j S^i)-{i\over2}
	({\pd S^i\over\pd p_j}-{\pd S^j\over\pd p_i}). \nonumber
\eqlab{ctterms}
}
At order $\leq\!\!-3$, however, the homomorphism equation \eqref{rhohom} has
no solutions for an Ansatz of the form \eqref{rhozero}--\eqref{rhotwo}. 
Thus, the best we can achieve is a representation of the algebra 
\eqref{mapext} on the quotient $\glopsdo/\idl$, where $\idl$ is the ideal
of PSDOs of order $\leq-3$. 

On the other hand, it is easy to see from the definition \eqref{trad} that
$\crt$ vanishes if one of its arguments lies in $\idl$.
Since, furthermore, a non-trivial calculation shows that    
\eq{
\ohom^*\crt(X,Y)\equiv\crt(\ohom(X),\ohom(Y))=\omega_3(X,Y)
}
when $\ohom$ is given by \eqref{ctterms}, $\ohom$ generates the correct
central term. Hence, we have obtained a
homomorphism $\ohom:\oalgth\rightarrow\widetilde{\glopsdo}/\idl$, where 
the central extension $\widetilde{\glopsdo}$ of $\glopsdo$ is defined by the 
twisted Radul cocycle $\crt$.

We now wish to second-quantize this representation. 
To this end, first note that the Fock space operators
\eq{
\wh{\ohom(X)}=\sum_{m,n}\ohom(X)_{mn}:\crm\ann:
\eqlab{tsec}
}
obey the commutation relations of $\gloh$ with the ill-defined 
Schwinger term $c_1(\ohom(X),\ohom(Y))$. Instead we define
\eq{
\dwh{\ohom(X)} = \wh{\ohom(X)}+\lambda(\ohom(X)),
\eqlab{dtsec}
}    
where $\lambda$ is the one-cochain entering in \Eqref{cohorel} \cite{CFNW-I}. 
{}From the fact that $\delta\lambda(S,S')=\lambda([S,S'])$ it
follows that the operators \eqref{dtsec} satisfy
\eq{
[\dwh{\ohom(X)},\dwh{\ohom(Y)}] \!=\! \dwh{[\ohom(X),\ohom(Y)]}
  +\crt(\ohom(X),\ohom(Y)),  
\eqlab{dtalg}}
where the Schwinger term is now well defined.

However, since
\eq{
\dwh{\ohom(\ham)}\fvac=\lambda(\ohom(\ham))\fvac,
}
the original Fock vacuum is not a HW state
for a representation of the algebra \eqref{dtalg} on $\fock$. 
Unfortunately, we have not been able to find such a vacuum state. 
In addition, we need a way to characterize Fock space states that are
annihilated by the ideal $\idl$ in order to be able to divide out this 
ideal from the representation.

\section{Conclusions and comments}
\seclab{concl}

Although we have obtained a potentially interesting realization of a 
higher-dimensional loop algebra as a current algebra 
on a fermionic Fock space, it appears difficult to overcome the
remaining obstacles to the construction of a unitary HW representation.
On the other hand, it is fair to say that had our construction worked
completely, it would have been a truly remarkable result.  

In any case, the search for this representation has lead to some nice
mathematical results \cite{CFNW-II}, namely the proof of the cohomological 
equivalence on the algebra of second-quantizable PSDOs between the Lundberg 
(a.k.a.~Kac--Peterson) cocycle defining the universal central extension of 
$\glo$ and a twisted version of the Radul cocycle. 
We also showed in ref.~\cite{CFNW-II} that, in arbitrary number of dimensions
$d$, the Radul cocycle may for two arbitrary PSDOs be written as the phase 
space integral of the order-$(-d)$ part of the star commutator of their 
symbols.

\section*{Acknowledgments}
The author wishes to thank M. Cederwall, G. Ferretti and B.E.W. Nilsson for
a pleasant collaboration.

\end{document}